\documentstyle[floats, preprint,tighten,epsf,prd,aps]{revtex}
\def\be{\begin{equation}}
\def\ee{\end{equation}}

\def\bea{\begin{eqnarray}}
\def\eea{\end{eqnarray}}
\def\bml{\begin{mathletters}}
\def\blea{\begin{mathletters}\begin{eqnarray}}
\def\elea{\end{eqnarray}\end{mathletters}}

\def\pprior{P_{\rm prior}}
\def\tpast{t_{\rm past}}
\def\tfuture{t_{\rm future}}
\def\Npast{N_{\rm past}}

\begin{document}
\draft
\title{The doomsday argument and the number of possible observers}
\author{Ken D. Olum\footnote{Email address: {\tt kdo@alum.mit.edu}}}
\address {Institute of Cosmology, 
Department of Physics and Astronomy, 
Tufts University, 
Medford, Massachusetts 02155}
\date{October 2000}

\maketitle
\begin{abstract}
If the human race comes to an end relatively shortly, then we have
been born at a fairly typical time in history of humanity.  On the
other hand, if humanity lasts for much longer and trillions of people
eventually exist, then we have been born in the first surprisingly
tiny fraction of all people.  According to the Doomsday Argument of
Carter, Leslie, Gott, and Nielsen, this means that the chance of a
disaster which would obliterate humanity is much larger than usually
thought.  Here I argue that treating possible observers in the same
way as those who actually exist avoids this conclusion.  Under this
treatment, it is more likely to exist at all in a race which is
long-lived, as originally discussed by Dieks, and this cancels the
Doomsday Argument, so that the chance of a disaster is only what one
would ordinarily estimate.  Treating possible and actual observers
alike also allows sensible anthropic predictions from quantum
cosmology, which would otherwise depend on one's interpretation of
quantum mechanics.
\end{abstract}

\pacs{98.80.Hw	% mathematical aspects of cosmology; quantum cosmology
      02.25.Cw	% Probability theory
      01.75.+m	% science and society
}

\section{Introduction}
The Doomsday Argument was introduced by Carter \cite{Carter:original}
and Leslie \cite{Leslie:1989}\footnote{Leslie has written extensively
on this subject; his ideas are collected and summarized in
\cite{Leslie:book}}, and independently in somewhat different form by
Gott \cite{Gott:doomsday} and also by Nielsen \cite{Nielsen:doomsday}.
The general argument in the Carter-Leslie form runs as follows: There
is some possibility that the human race will last for a very long time
and grow to huge numbers.  In that case, we have been born in the
first tiny fraction of all humans, which would be very surprising.  On
the other hand, there is also some possibility that the human race will
die out before too long due to some disaster (nuclear war, asteroid
impact, etc.).  In that case, we have been born roughly in the middle
of all humans.  (About 10\% of humans ever born are alive today, so
even if the human race were to end tomorrow, the average person alive
today would just be in the last 10\%.) This, claim Carter and Leslie, gives us
reason to increase our estimates that the race will end before too
long, making us typical, rather than going on a very long time, making
us unusual.

The underlying idea is formalized by Bostrom \cite{Bostrom:thesis} as the
``Self-Sampling Assumption'':
\begin{quote}
Every observer should reason as if they were a random sample drawn
from the set of all observers.
\end{quote}
This is essentially Vilenkin's ``principle of mediocrity''
\cite{Vilenkin:1995ua} applied to individual observers rather than
civilizations.  To derive the doomsday conclusion, according to Leslie
and Bostrom, one proceeds as follows \cite{Bostrom:thesis}.  Something
like 60 billion people have been born so far.  Suppose for simplicity
that there are only two possibilities: Either the race will die out
soon, so that the total number of humans ever to be born is 200
billion, or it will last much longer so that the total number is 200
trillion.  We will call the hypotheses of a short- or long-lived race
$S$ and $L$ respectively.  Suppose that before you take into account
the doomsday argument, you think that the chances of these two
alternatives are $\pprior (S)$ and $\pprior (L) = 1-\pprior (S)$.  Now you
take into account the fact that you are the $N$th human to be born,
with $N\sim 6\times 10^{10}$.  The probabilities of $S$ and $L$ should
be multiplied by the chance that your birth rank would have been $N$ in
each case, which we will denote $P (N|a)$, where $a$ ranges over $S$
and $L$.  Thus the chance of each alternative, taking into account
$N$, is
\be
P (a|N)\propto P (N|a)\pprior (a)\,.
\ee
Including the normalization factor, you find that the
probability that the human race will be short-lived is now
\be\label{eqn:Bayes} 
P (S|N) ={P (N|S) \pprior (S)\over P (N|S) \pprior (S)
+P (N|L) \pprior (L)}\,,
\ee
which is just Bayes's  Rule.
The chance to have a particular birth rank is inversely
proportional to the total number of humans to ever exist, so $P
(N|S) = 1000 P (N|L)$ and
\be
P (S|N) ={\pprior (S)\over \pprior (S) +10^{-3} \pprior (L)}
\ee
Unless you started with $\pprior (S)\sim 10^{-3}$ or less, you will  find that
the chance of a short-lived race is nearly 1.

Gott \cite{Gott:doomsday} makes a similar claim about the length of
time for which the human race will exist, based on the ``delta $t$
argument''.  If a phenomenon exists for certain period of time $T$,
then we should expect ourselves to be observing it at a random time
during its life.  Thus, for example, the chance that we see the
phenomenon in the first or last 2.5\% of its life is 0.05.  Gott
argues from this that if we know the amount of time $\tpast$ for which
the phenomenon has existed, then we can conclude that there is a 95\%
chance that the total lifetime will be such as to make
\be
0.025T <\tpast < 0.975T\,.
\ee
We can rewrite this inequality in terms of $\tfuture = T-\tpast$,
\be
\tpast/39 <\tfuture < 39\tpast\,.
\ee
Applying this to the human species, Gott uses $\tpast\approx$ 200,000
years to conclude that there is a 95\% probability that the future
lifetime of our species will be between 5,100 and $7.8\times 10^6$
years.
 
Many counterarguments can be raised against the doomsday argument, but
I want to concentrate here on a single one, as follows.  In the
scenario where the human race is very long-lived and there are many
humans altogether, there is a greater ``chance to exist at all'' than in the
scenario where the race is soon to die out.
Thus our prior probability should first be multiplied by the number of
people in each scenario.  We can write
\be
P (a|I) \propto N_{\rm total} (a) \pprior (a)
\ee
where $N_{\rm total} (a)$ is the total number of observers in case
$a$, and $P (a|I)$ is the probability of case $a$ given that I
exist to observe it.  Then
\be\label{eqn:null}
P (a|N)\propto P (N|a) P (a|I)\propto P (N|a) N_{\rm total} (a)\pprior
(a) \propto\pprior (a) \,.
\ee
Thus the increased chance of finding oneself in a long-lived race,
because it contains more observers, exactly cancels the decreased
chance of finding oneself with a particular $N$ in the long-lived
race.  The chance of the race dying out quickly is thus the prior
chance of such an event, whatever one computes that to be based on one's
estimation of the various possible disasters.  The doomsday argument
does not modify the conclusion.

The idea that one is more likely to find oneself in the long-lived
race is called the Self-Indication Assumption by Bostrom.  It was
first discussed by Dieks \cite{Dieks:sia}, and has since been
criticized by Leslie \cite{Leslie:book} and Bostrom
\cite{Bostrom:thesis} and defended by Kopf, Krtous and
Page \cite{Page:sia} and by Bartha and Hitchcock \cite{Bartha:sia}.  I
will give several arguments in favor and attempt to answer Bostrom's
and Leslie's objections.  I will also analyze Gott's argument
specifically, and address the issue of why, if it is wrong, it seems
to have had some success predicting the lifetimes of various
phenomena.

\section{Possible vs.\ existing}
\subsection{God's Coin Toss}

The crux of the matter can be described by a ``God's Coin Toss''
experiment \cite{Leslie:book,Bostrom:thesis}.  Suppose that God
tosses a fair coin.  If it comes up heads, he creates ten people, each
in their own room.  If tails, he creates one thousand people, each in
their own room.  The rooms are numbered 1-10 or 1-1000.  The people
cannot see or communicate with the other rooms.  Suppose that you know
all this, and you discover that you are in one of the first ten rooms.
How should you reason that the coin fell?

Leslie and Bostrom argue as follows.  Before you look at your room
number, you should think that since the coin was fair the chance of
heads was 1/2.  Now if the coin was heads, then of course you would be
in one of the first ten rooms.  However, if the coin was tails, the
chance to be in one of the first ten rooms is 1/100.  Thus, according
to Eq.\ (\ref{eqn:Bayes}), you should now believe that the coin was
heads with probability 0.99.

The alternative argument runs as follows.  Before you look at your
room number, you should think that the probability of heads is 0.99.
There are one thousand possible people who would be right with that
belief, whereas only ten would be right with the belief in heads.
When you look at your room number, you should then update your
probabilities using Eq.\ (\ref{eqn:Bayes}).  The result is that in the
end you think the chance is 1/2 that the coin was heads.  Another way
to say the same thing is that there are ten ways to have the coin
heads and you in a room in the first ten, and ten ways to have the
coin tails and you in a room in the first ten, and thus the chances
for heads and tails are equal.

The difference here hinges on whether one considers possible people in
the same ways that one considers actual people.  If instead of
flipping a coin, God creates both sets of rooms, then Leslie and
Bostrom and I all agree that you should think it much more probable
that you are in the large set before you look at your room number, and
equally probable afterward.  Treating the two possibilities in the
same way as two sets of actual observers implies the Self-Indication
Assumption: the existence of a large number of observers in a possible
universe increases the chance to find oneself in that universe.

I will argue below that the equal treatment of possible and actual
observers is correct.

\subsection{Improving the experiment}

It is possible to produce modified versions of this thought experiment
which will avoid any disagreement, as follows.\footnote{ Similar
thought experiments have been discussed by Bartha and Hitchcock
\cite{Bartha:sia} and by Bostrom \cite{Bostrom:thesis}.}  Imagine that
you are one of a very large number of experimental subjects who have
been gathered, in case of need, into an experimental pool.  Each
subject is in a separate waiting room and cannot communicate with the
others.  First the experiment will be described to you, and then it
will be performed.  The experiment will have one of the following two
designs.

\newtheorem{protocol}{Protocol}
\begin{protocol}[random]
The experimenter will flip a fair coin.  If the coin
lands heads, she will get ten subjects, chosen randomly from the pool,
and put them in rooms numbered 1-10.  If the coin lands tails, she
will do the same with one thousand subjects in rooms numbered
1-1000.
\end{protocol}

\begin{protocol}[guaranteed]
The experimenter will flip a fair coin.  If the coin
lands heads, she will get you and nine other subjects, and put you
randomly into rooms numbered 1-10.  If the coin lands tails, she will
get you and 999 other subjects and put you randomly into rooms
numbered 1-1000.
\end{protocol}

How should you rate the probability of the outcomes of the coin flip,
before and after learning that your room number is in the first ten?
I think we can all agree that in protocol 1, before looking at your
room number you should expect that the coin was tails with probability
0.99, because it is 100 times more likely in that case that you would
have been chosen for the experiment at all.  Then when you have
learned your room number, you should think that the chance of heads
was 1/2.

In protocol 2, since you knew that you would participate, you don't
learn anything more when the experiment begins, so you should think
the chance of heads was 1/2.  After learning that you are in one of
the first ten rooms, you should think that the chance for heads is now
0.99, in accordance with Eq.\ (\ref{eqn:Bayes}). 

The question, then, is which of these scenarios is like the God's Coin
Toss experiment.  If God's Coin Toss is like protocol 1, then the
chance of there being many people in this case is equal to the
chance of there being few, and the Doomsday Argument is wrong.  If
God's Coin Toss is like protocol 2, then it is much more likely that
there are few people, and the Doomsday Argument is correct.  I believe
that it is the first case, and thus the Doomsday Argument is wrong.
The following section gives several arguments in support of this position.

\section{Arguments}

\subsection{Asymmetry}

Protocol 1 is symmetrical with respect to all participants.  That
is, each person sitting in their room, before looking at the number,
can reason the same way.  Protocol 2, however, does not have this
property.  You, as one of the originally chosen subjects, can reason
as above.  However, at most 9 other people can reason in this way.  If
there are one thousand subjects, the rest of them must have started
with different initial information.  The God's Coin Toss example is
symmetrical with respect to all participants, and so is like
protocol 1.

Note that it does not help to argue that you exist, and thus that you
must be one of those chosen for the experiment.  This is merely like
observing, as you sit in your numbered room, that you {\em have been
chosen} to participate.  It applies equally in the case of protocol
1, and does not change the argument there.  To claim that the God's
coin toss (or the real world) is like protocol 2, you have to claim
not only that you exist but that you exist {\em necessarily}, i.e.,
that God was required to choose {\em you} as one of the people to
create, regardless of the outcome of the coin flip.  You must also
think that in case the coin landed tails, there would be lots of other
people who were created but whose creation was not necessary.

\subsection{Discontinuity}
The Leslie-Bostrom analysis of God's Coin Toss has a strange
discontinuity in the way in which your judgment of the chance of heads
depends on the number of people created in that case.  Consider the set
of cases where on tails God creates $10^{10}$ people, and on heads he
creates some number $N$ between 0 and $10^{10}$, and the particular
value of $N$ is known to you.  If $N = 0$, then you can say with
certainty that the coin fell tails, because otherwise you would not
exist.  If we treat the $N$ and the $10^{10}$ possible people the same
way we would treat actual groups of people, then the probability we
assign to heads smoothly approaches 0 as $N\to 0$.  But by Leslie and
Bostrom's analysis, this probability is always 1/2 as long as $N\ge1$,
but drops suddenly to 0 when $N = 0$.  

\subsection{Dependence on the nature of probability}
\label{sec:qm-interpretation}
There are different kinds of probability.  One kind is based on
ignorance, such as the case of the coin which has already been flipped
but not yet examined.  A similar kind of probability is the case of an
event which has already been determined, but not yet occurred, and
whose occurrence is too complicated to compute.  An example is the
falling of the ball into a slot on a roulette wheel after you make a
last-minute bet (but see \cite{Eudaemonic}).  Another example is the
pseudo-random numbers generated by a computer.  In the deterministic
Newtonian worldview, all probabilities are of one of these types.

In contrast to this view, one can have classical indeterminism, in
which outcomes of (some) events are determined truly by chance.  One
can also have quantum mechanical indeterminism, but
that depends on one's interpretation of quantum mechanics.  In the
interpretation which just takes the wave function as fundamental with
no collapse process, there is no indeterminism.  There is also no
indeterminism in the multiple worlds interpretation.

The multiple worlds interpretation is particularly problematic here.
If one imagines that God's Coin Toss is really a quantum mechanical
event, then in the literal multiple worlds interpretation, there are
two sets of actually existing observers, one in ten rooms and one in a
thousand rooms.  In such a case, advocates of the doomsday argument
agree that it does not apply.  The claim that the likelihood
of disaster depends on the interpretation of quantum mechanics
is a strange one, since physical processes (such as those that might
lead to destruction) should not depend on a choice of quantum
mechanics interpretations.

\subsection{Dependence on the content of causally disconnected
regions} At least as argued by Bostrom \cite[page 123]{Bostrom:thesis}, the doomsday argument depends on whether or not
we are the only intelligent species in the universe.  He claims that
if there are lots of extraterrestrial civilizations, of varying
duration, then we should expect to find ourselves in one of the
long-lived civilizations, thus canceling the doomsday argument.  This
means that how one should estimate one's future prospects depends on
the existence of extraterrestrials, even if they are in
causally-disconnected regions of the universe.  Again, this is very
strange, because one would think that conditions in places that can
have no communication with us should not affect our future.

\subsection{Reference class problems}
The {\em reference class} is the set of all observers, which one needs
(explicitly or implicitly) in order to say things like ``one expects to
find oneself randomly situated among all observers''.  There are
many problems concerning the definition of this class.  However, there
is a special problem related to different treatment of possible and
actual observers.  Suppose that you know that God created one set of
ten rooms with ten humans and another set of a thousand rooms with
humans in the first ten and chimpanzees in the remainder.  Finding
yourself a human in one of rooms 1-10, you can conclude that it is
equally likely that you are in the set of just ten or the first ten in
the thousand. In this case, it is not necessary to discuss the mental
capacities of chimpanzees.

But now suppose that God flips the coin and creates {\em either} just
the ten humans (if heads), {\em or} the ten humans and 990 chimpanzees
(if tails).  I would argue that finding yourself a human in one of
rooms 1-10, you should conclude that the chance that the coin fell
heads is 1/2.  But if you believe the analysis of Leslie and Bostrom
\cite[page 77]{Bostrom:thesis}, then the result depends
crucially on whether you could have been a chimpanzee.  If chimpanzees
are in the reference class, then the argument is the same as the
original God's Coin Toss, and it is nearly certain that the coin fell
heads.  On the other hand, if chimpanzees are excluded from the
reference class, perhaps because they can't understand philosophical
arguments, then you are necessarily in one of rooms 1-10, and so the
chance of heads is still 1/2.  Thus the different treatment of
possible observers has led to a situation where you need to know
something about the intellectual capacity of chimpanzees, {\em even
though you already know you aren't one}.

\subsection{Unreasonable predictive powers}

Although it may seem to beg the question, one can argue that the
existence of the doomsday argument may itself be a reason not to
believe that possible and actual observers should be treated
differently.  The point is not the undesirability of the doomsday
conclusion, but rather that it does not seem that one should be able
to infer such conclusions about the future by looking only at the
past.  The probability of a disaster should be just the probability
that it will occur given the pre-existing conditions.

One might imagine the case of a gambler who is about to throw a set of
fair dice.  If she wins she will spend her winnings on some action
that will increase the eventual number of people, such as the funding
of space colonies.  Should she then think that her chance for the dice
to fall favorably is reduced below the normal statistical probability?

This problem can be further strengthened, as follows.

\subsection{Paranormal and backward causation}
Bostrom \cite[chapter 8]{Bostrom:thesis} points out that if one accepts
the doomsday argument one must also accept a number of very strange
similar arguments.  For example, suppose that there is some kind of a
natural happening that we have no control over but nevertheless wish
to avert.  For example, suppose that we learn that a nearby star has a
90\% chance of becoming a supernova, causing significant destruction
on earth but not killing everyone.  Now we make the plan (and make
sure that it will be carried out) that if the supernova occurs we will
start an aggressive program of space colonization, leading to a huge
increase in the number of people that will eventually exist, while
otherwise we will not.  Now the same doomsday argument that says that
the human race is likely to end soon tells us that the supernova is
not likely to occur.  If it did occur, we would then be in the first
tiny fraction of humanity.

Not only does this seem to allow us to affect things over which we
should not be able to have any control, but it even works backward in
time.  By exactly the same argument, even if the supernova has or has
not already occurred in the past, and its effects have not reached us,
we can change the chance of its {\em having occurred} by the above
procedure.

Obviously this kind of paranormal and backward causation is
ridiculous.  Bostrom says that it is not as bad as it seems, but to
do so he has to resort to some rather strange argumentation
including the claim that given some action A and some consequence C,
one can consistently believe ``If we do A then we will have brought
about C'' and ``If we don't do A then the counterfactual 'Had we done A
then C would have occurred' is false''.  It would seem easier just to
say that the type of argumentation that gives us these paranormal
powers, and thus the Doomsday Argument as well, is simply incorrect.

We can perhaps understand this situation better by trying to construct
it using the experimental protocols above.  In the random assignment case,
there is no problem.  We get

\begin{protocol}[random assignment, late decision]  The experimenter will
first choose ten subjects randomly and put them in rooms 1-10.
Subject \#10 will then get to flip a coin and call the outcome.  If he
is correct, the experiment is over.  If not, the experimenter will
choose subjects randomly to fill the rest of the thousand rooms.
\end{protocol}

Does this give subject \#10 any special ability to influence the coin
flip?  No, because of the reasoning above with respect to protocol 1.
Finding yourself in the first 10 rooms in this case gives you no
special information about the coin flip.

To produce an analogy where the supernova argument would work is very
tricky.  In the case where the coin is indeterministic, it cannot be
done.  The experimenter must first fill rooms 1-10 without knowing the
flip outcome.  Thus she doesn't know whether to put you (the
privileged observer) into one of these rooms or not, so she cannot
create situation where you have an equal chance of being in any
of the thousand rooms, in the case of incorrect prediction.  I suspect
that this issue is at the heart of Leslie's \cite{Leslie:book} claim
that the doomsday argument depends on the nonexistence of ``radical
indeterminism''.

In the case where there is determinism, then perhaps the experimenter
can arrange the experiment as desired by advance knowledge of how the
coin will fall and how subject \#10 will call it, giving

\begin{protocol}[guarantee, late decision]  The experimenter will choose
ten subjects and put them in rooms 1-10.  Subject \#10 will then flip
a coin and call the outcome.  If he will be correct, then the
experimenter will make you one of the first ten subjects at random.
If subject \#10 will be incorrect, then the experimenter will fill the
remainder of the thousand rooms, and will assign you to one of the
entire thousand rooms at random.
\end{protocol}

Now can you infer that subject \#10 has special powers to predict the
coin flip?  Indeed you can.  Your chance to have been in one of the first 10
rooms is very small if he's going to call it wrong, while it is
guaranteed if he's going to call it right.  On the other hand, one can
see that this is really just that the experimenter has given you some
special knowledge about the outcome of her amazing predictive
abilities.  Furthermore, she cannot have given this information to
everyone, because all ``guarantee'' protocols treat some of the
observer specially.  I don't think the real world where we are
trying to avert some disaster has much in common with this model.

\subsection{Nonrepeatability}
Suppose that God flips his coin many times and creates many batches
of people, some in sets of ten and some in sets of a thousand.  Then
even Leslie's argument doesn't yield a doomsday prediction.  All sequences
of coin flips are equally probable, in this interpretation, but the
great majority of the time there are many small sets and many large sets.  Thus before looking at your room number,
you expect to be in a large set.  After finding that you are in the
first ten, you now reduce your estimate of the coin flip relevant to
you to nearly equal probabilities of heads and tails.  The same result
could be seen in a repeated version of the ``guarantee'' protocol in
which the experiment is done many times and the guarantee is that
you will be one of the subjects in one of the runs, with 
equal probability to be any one of those subjects.

Thus Leslie's argumentation depends on there being only a single
universe.  He agrees with this and says ``In cases like this we must
reject the intuition\ldots that to estimate probabilities we ought to
ask what bets would maximize winnings when the experiment was repeated
infinitely many times'' \cite[page  228]{Leslie:book}.  This does not seem
to bother Leslie, but it is a strange claim.  In some theories of
probability, probability just means the bet that would maximize
winnings, but even if one doesn't accept this as a definition, it
still seems clear there is something wrong with a system for
computing probabilities that yields odds vastly different from how a
bettor should bet.

\section{Gott's analysis}

\subsection{Problems}
Gott's argument, discussed in the introduction, suffers from two
important errors pointed out in a letter by Buch \cite{Buch:letter}.
They are discussed in detail by Caves \cite{Caves:Gott} and by Bostrom
\cite{Bostrom:thesis}, so I cover them somewhat quickly here.  First of
all, Gott \cite{Gott:doomsday} makes a simple omission: he does not
include the prior probabilities for the various lifetimes.  
Gott corrected this omission in \cite{Gott:letter} and
\cite{Gott:proceedings} by saying that we should start with the ``vague
Bayesian prior'' or ``Jeffreys prior'' \cite{Jeffreys:book},
\be\label{eqn:Gottprior}
\pprior (T)\propto T^{-1}\,.  
\ee 
This means that the chance for $T$ to be in any logarithmic
interval is the same.

Once we have the prior probability, Gott claims we can determine the
probability for various lifetimes according to Bayes's Rule as
\be\label{eqn:GottBayes}
P (T|\tpast) \propto P (\tpast|T)\pprior (T)\,.
\ee
Here $P (\tpast|T)$ is the chance to measure $\tpast$ given that the
actual lifetime is $T$, which is
\be\label{eqn:past}
P (\tpast|T) =\cases{1/T & if $\tpast < T$\cr 0 & otherwise.}
\ee
Putting Eqs.\ (\ref{eqn:Gottprior}) and (\ref{eqn:past}) into Eq.\
(\ref{eqn:GottBayes}) gives
\be\label{eqn:Gott2} 
P (T|\tpast) \propto 1/T^{-2}\qquad{\rm if}\, T >\tpast
\ee 
which reproduces Gott's predictions.

Unfortunately, there are still two problems.  The first is that Gott's
prior is not reasonable for any ordinary phenomenon.  For one thing,
it cannot be normalized, and even if one doesn't consider this a
technical difficulty, it means that the chance of $T$ being in any
given finite interval is zero.  Even if we adopt Gott's suggestion to
establish cutoffs at some tiny and some huge lifetime
(e.g., \cite{Gott:letter} recommends the upper limit $10^{5,000,000}$
years), the prior chance of any reasonably sized interval is
infinitesimal.  For example, Gott applied his principle to the Berlin
Wall, built in 1961 and observed by Gott in 1969.  Would it be
reasonable to use the ``vague prior'' here?  To check this we can ask
what would be a reasonable expectation for the lifetime of the wall at
the time that it was built, and thus when no ``past lifetime''
information was available.  How would you have estimated the chance
that it would last more than a year and less than 100 years?  More
than a year and less than 100,000 years?  If you think that either
chance is nonzero, then you don't believe in the vague prior.  (In the
case with the cutoff, there would be 5,000,000 factors of 10 in the
range of possible durations, so the chance to be between 1 and $10^5 $
years is $10^{-6}$.)

The second problem \cite{Buch:letter,Caves:Gott} is that Gott has neglected the fact that a
long-lived phenomenon is more likely to be presently ongoing, and thus
to be observed, than one which is short-lived.  Thus if $\pprior (T)$ is
the probability that a phenomenon of the class under consideration
chosen randomly from among all such phenomenon ever to exist has
lifetime $T$, then the probability that a phenomenon chosen from all
those currently existing has this lifetime is
\be\label{eqn:anthropicfactor}
P_{\hbox{\scriptsize currently existing}} (T)\propto T\pprior (T)\,.
\ee
Following \cite{Buch:letter}, we call this the anthropic factor.
Its effect is to cancel the factor due to $P
(\tpast|T)$ to give
\be\label{eqn:Gottnull}
P (T|\tpast) \propto \cases{\pprior (T) & if $T > \tpast$\cr 0 & otherwise.}
\ee
Thus we do not learn anything new from knowing the past lifetime, other
than that the total lifetime must be at least as large as what we have
observed.  Our judgment of the longevity of the phenomenon is still
what we get from the prior probability, i.e. from our analysis of
other information we may have about its lifetime.

The anthropic factor of Eq.\ (\ref{eqn:anthropicfactor}) is very much
like the argument of previous sections that an observer is more
likely to observe a world with many observers than one with few, and
the cancelation that leads to Eq.\ (\ref{eqn:Gottnull}) is very much
like the cancelation that leads to Eq.\ (\ref{eqn:null}).  In both
cases, when one takes the relevant principle into account, one finds
that the conditional probability based on one's observations is just
the prior probability.

However, in the present case there is no need to resort to complex
argumentation to justify the use of Eq.\ (\ref{eqn:anthropicfactor}).
For a simple example, we can consider the case of radioactive decay
\cite{Buch:letter}, where the prior is known.  Consider nuclei of some
radioactive element with lifetime $\tau$.  Each such nucleus will live
for sometime $T$ after the time at which it is created, and the
distribution of $T$ is
\be
\pprior (T)\propto e^{-T/\tau}\,.
\ee
Now suppose that we find such a nucleus lying about our laboratory,
and somehow we're able to trace its history and learn its $\tpast$.
Now by Gott's argument, we should think that its total lifetime is
distributed as
\be
P (T)\propto{1\over T} e^{-T/\tau}\qquad{\rm if}\, T >\tpast\,,
\ee
which is not correct.  Including the anthropic factor
gives the correct answer
\be
P (T)\propto e^{-T/\tau}\qquad{\rm if}\, T >\tpast\,.
\ee
Thus in an elementary case, we see that it is
correct to include the extra factor, and thus that no ``extra''
information can be derived from considering the past lifetime.

\subsection{Successes}

Gott made many predictions by applying his principle, and many of them
appeared to be confirmed.  If his principle was wrong, why did his
predictions turn out so well?  One reason is that using an
unreasonable prior probability and not including the anthropic factor
cancel each other to some degree.  It's hard to know what would be a
reasonable prior probability for examples like the Berlin Wall, but
Gott also studied the lifetimes of Broadway and off-Broadway plays and
musicals, and in this case we have a large dataset and can do
statistics.

\begin{figure}[bt]
\begin{center}
\leavevmode\epsfbox{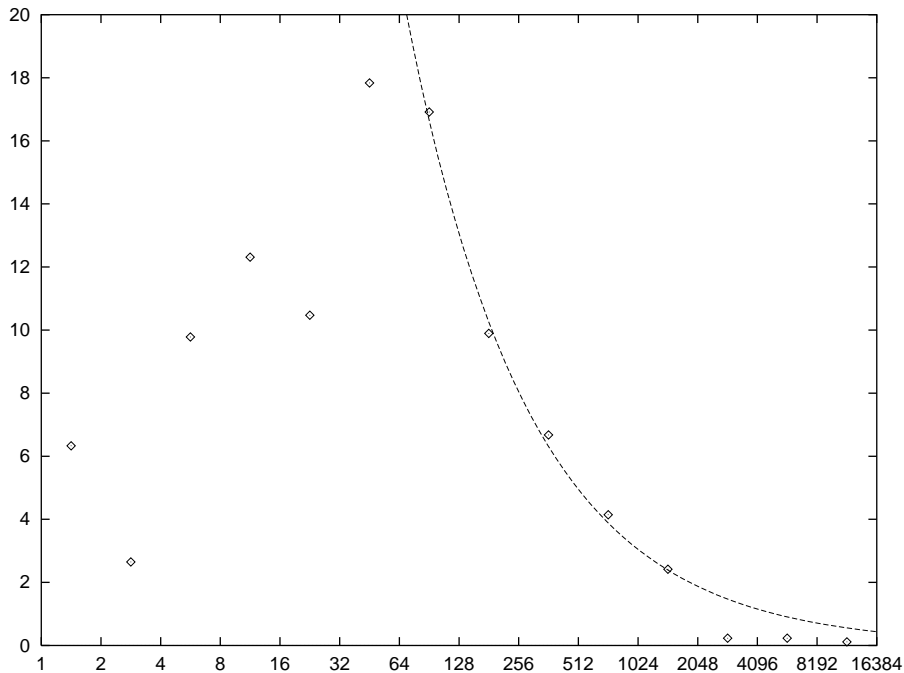}
\end{center}
\caption{Percentage of shows with running times more than or equal to
the beginning of each interval and less than the end.  The dashed line
is a power-law fit to the data points after the peak, $P = 390T^{-0.7}$.}
\label{fig:shows}
\end{figure}

I analyzed all the Broadway shows listed in the Internet Theatre Database
(www.theatredb.com) that opened between 6/1/70 and 5/30/90 to
determine their running time.\footnote{I didn't include off-Broadway
shows, because they were not uniformly listed in this database, and I
didn't include any after 1990, in order to give them time to close.  I
only listed shows that had closed and whose opening and closing dates
were recorded in this database.  About 5\% of the shows did not have
this data, but they did not seem to be a biased set.}
The data are shown in Fig.\ \ref{fig:shows}.
The data points are the percentage of shows whose running time (number of
days counting both opening and closing days) is within the given
number of days.  Thus the first data point is the number of shows
running only one day, the next is those running two or three days, and
so on.

It is clear from Fig.\ \ref{fig:shows} that the actual distribution
does not fit Gott's prior very well at all.  The ``vague prior'', Eq.\
(\ref{eqn:Gottprior}), would give equal numbers of events in each
interval, and thus a flat distribution in Fig.\ \ref{fig:shows}.
Instead, there is a peak around 50--100 days, and a sharp decline
after that.

However, for durations after the peak, and excluding a few very long
duration shows, the number of shows in a logarithmic interval is well
fit by the curve $\pprior (T) dT\propto T^{-0.7} d\ln T\propto
T^{-1.7} dT $, shown by the dashed line in Fig.\ \ref{fig:shows}.
This is close to $\pprior (T) \propto T^{-2}$, which 
yields Eq.\ (\ref{eqn:Gott2}) once one
properly takes account of the anthropic factor, and thus reproduces
Gott's calculations \cite{Caves:Gott}.  This explains some of
Gott's successes in predicting the longevity of long-lived shows.

What about shows with short $\tpast$?  It's clear that Gott
underpredicts their duration.  For example, Gott gives a show which
has been open 44 days a 97.5\% chance of closing within $4\times 39 = 156$
days, but from the 20 years of data described above, this chance is
only 78\%.  However, very few of Gott's sample had been showing for
such low periods of time, so the number of opportunities for him to
make such an error was small.

What about the future of humanity?  Gott claims that since
$\Npast\sim7\times 10^{10}$ people have been born so far, the number of
humans yet to be born is between $1.8\times 10^9$ and $2.7\times
10^{12}$, and thus the current birthrate is unlikely to continue for
more than 19,000 years.\footnote{This argument is more pessimistic
than that of the introduction, because it considers the number of
people rather than the duration of civilization in years.}  This
argument, however, neglects the ``anthropic'' factor, and uses the
``vague prior''.  If we do the calculation correctly, we just
reproduce the prior probability, and so the issue is to make our best
estimation of our probable survival.  One could debate this estimate at length,
but it is clear that it is not the ``vague prior'', since there is
obviously some chance of our species perishing in the near future.  In
fact, it seems that the previous lifetime of our species is
essentially irrelevant to many of the possible causes of our
destruction.  It doesn't really matter how long it took and how many
people lived and died before we achieved the technology to destroy
ourselves.  The danger is at the present time when we have the ability
to destroy our species but lack good safeguards that would prevent
us from doing so.

Moreover, it seems as though a sensible estimation of our chances is
bimodal.  There is a nonzero chance that we will become extinct or lose our
technological abilities in the relatively near future.  But if this
does not happen, if it is likely that we will colonize the galaxy, and then
our chances of surviving a very long time are quite high.  Thus a
sensible distribution for the lifetime of humanity or the number of
people who will ever be born is not smooth or monotonic.

\section{Counterarguments}

\subsection{Confirming evidence}
Leslie \cite[page 226]{Leslie:book} gives the following example about the
probability of theories that yield different numbers of observers.
Marochnik \cite{Marochnik} suggested a theory in which
planets (and thus observers) occur only around stars near corotation
(the distance from the center of the galaxy at which the individual
stars orbit with the angular velocity of the spiral arm pattern).
Suppose (to strengthen the argument) that this suggestion was made
before our Sun's position was known, and then it was discovered that
it was quite close to corotation.  Surely this should be viewed as
important support for Marochnik's theory.

On the other hand, Leslie argues, if we give greater probability to
universes with more observers (the ``Self-Indication Assumption'' ---
SIA), the chance of being here at corotation is no larger in this
theory than in one in which planets were common throughout the galaxy.
Thus the theory that planets are everywhere and the theory that
planets are only here give the same probability for us to be where we
are.  It appears that in fact the evidence does not lead us to
prefer the Marochnik theory.  Leslie rejects this consequence and
claims that it shows that SIA is wrong.

What has happened here?  Is it really true that accepting SIA means
that we have no support for Marochnik's theory, even though it has
successfully predicted our position in the galaxy?
That's not quite right.  However you look at it, you should agree that
you have a much better reason to believe this theory after the
observation than before.  The issue is how likely this theory was in
advance of observation.  If you feel that in advance of observation,
the theory was wildly unlikely, then you can feel even though the
evidence has made it much more likely, that it is still not an
especially well-supported theory.  The reason to feel that it is a
priori unlikely, of course, is that it leads to a very small number of
observers, as opposed to the theory in which planets are common
everywhere.  (You could also feel that this theory is unlikely, in
advance of observation, because of the possibility that it will
immediately be ruled out by finding that we are in a place where no
planets should have been.)

Let's consider instead, for a moment, a problem about different
galaxies instead of different theories.  Suppose that some galaxies
have planets around every star, while others have planets only in a
special place.  Once we have learned this, we should think it highly
likely that our galaxy is one with planets everywhere.  However when
we later find that we ourselves live in the special place where all
galaxies have planets, then we should feel no preference about which
type of galaxy we have.  It would not be right to say that it's more
likely that we are in a few-planet galaxy because that would ``explain'' or
``correctly predict'' our position in our galaxy.  Half the planets in
our position are in few-planet galaxies, so we should think the chance
that we are one of those is 1/2.

Now consider an intermediate case where the number of planets depends
on some other parameter of cosmology.  For example,
consider a situation where you think that the hot dark matter (HDM)
and cold dark matter (CDM) theories of cosmology are equally likely.
Then suppose you find out that CDM would lead to planets around every
star, while HDM would lead to planets only in a special
place.\footnote{This is not supposed to be a real astrophysical theory,
but just an example to probe our reasoning about the credibility of
theories.} At this point, I think you should consider the CDM theory
much more likely to be correct, because it  predicts many observers,
while the HDM theory predicts few.  If now you find that we are in
fact in the special place where both theories predict planets, you
should return to thinking that the two theories are equally likely,
just as in the previous case.  The number of civilizations finding
themselves in the position which we find ourselves is the same in the
two theories, so we have no reason to prefer one over the other.

Now let us return to the Marochnik case.  Suppose that {\em before
considering the effect on the number of observers or measuring our
position in the galaxy} you think that the Marochnik theory and the
planets-everywhere theory are equally credible.  Then, I claim, when
you take account of the differing number of observers, you should
think that the planets-everywhere theory is far more likely, and when
you find our position in the galaxy is as Marochnik predicted, you
should go back to finding the theories equally likely. I realize that
it is quite counterintuitive that one should not believe this theory
when it has made a correct prediction, but I think the situation is as
in the model above, where the idea that we are in a few-planet galaxy
correctly ``predicts'' our location, but that does not make it more
probable that we are in fact in the few-planet galaxy.\footnote{This
raises the question of why we don't in fact find ourselves in a galaxy
teeming with intelligent species.  One possibility is that other
factors such as the amount of expansion of the universe during
inflation have dominated factors involving the number of intelligent
species per galaxy \cite{Vilenkin:1995ua}.}

\subsection{Bostrom's ``presumptuous philosopher''} 
Bostrom \cite[page 134]{Bostrom:thesis} gives a related
example.  Suppose that we are certain that one of two cosmological
theories is correct, but don't have a strong preference between them.
Both predict finite universes with contents similar to ours, but one
theory predicts a universe a trillion times larger than the other.
Physicists would like to do an experiment which will determine which
theory is the correct one, but the ``presumptuous philosopher'' explains
that this is unnecessary.  Since one theory has a trillion times more
observers than the other, we already know that that theory is a
trillion times more likely.

Is this in fact an argument against the Self-Indication Assumption, or
should we just accept that the presumptuous philosopher is correct?
As above, we can consider the case of a single agreed-upon theory that
includes a large number of universes, some a trillion times larger
than others, and we wanted to know which type of universe we were in.
Then the presumptuous philosopher would be right that the chance of
being in a small universe is infinitesimal.  It seems to me that, from
the arguments of this paper, the presumptuous philosopher is also
correct in the original scenario, as long as we feel that the
likelihoods of the two theories are roughly equal before one considers
the effect on the number of observers.

However, one should give at least some consideration to the idea that
a theory which involves a very large number is less likely to be
correct than one which does not.  For example, suppose I have a crazy
theory that each planet actually has $10^{10^{100}}$ copies of itself
on ``other planes''.  Suppose that I (as cranks often do) believe this
theory in spite of the fact that every reputable scientist thinks it
is garbage.  I could argue that my theory is very likely to be
correct, because the chance that every reputable scientist is
independently wrong is clearly more than 1 in $10^{10^{100}}$.  To
avoid this conclusion, one must say that the a priori chance that my
theory was right was less than 1 in $10^{10^{100}}$.  It seems hard to
have such fantastic confidence that a theory is wrong, but if we don't
allow that we will be prey to the argument above.

One might say that this really just shows the strange consequences of
accepting SIA.  However, similar scenarios exist without any
dependence on the number of observers.  For example, suppose a
stranger comes up to you with the claim that if you give him a dollar
today he will give you \$10 tomorrow.  Presumably you won't give him
the dollar, which shows (if you are maximizing your expectation) that
you think the chance he will come through as he says is less than
10\%.  On the other hand, it would be strange to claim that the chance
is less than one in a million, since sometimes people making statements
like this are honest.  At this level you might even consider the
possibility that your whole understanding of the world has one chance
in a million to be very wrong, and so you can't trust your expectation
that you won't be paid to this level.  Nevertheless, if the payoff is
raised to \$10 million, you still won't give the dollar, which shows
that now you think the chance for a payback is in fact less than 1 in
10 million.  In order to not have a proposed payback large enough to
deprive you of your money, you must think that the likelihood of
getting paid decreases at least inversely with the proposed payback.

Applying this to cosmology, it is possible that one should think that
a theory involving a very large universe is unlikely in proportion to
the size of the universe it proposes.  In this case, the presumptuous
philosopher is wrong, because the tiny a priori probability for the
theory with the larger universe cancels its larger number of
observers.  The theories must then compete on their merits.

\section{Quantum cosmology}

A theory of quantum cosmology gives an initial wave function for the
universe (although there is disagreement about what this wave function
should be \cite{Hartle:1983ai,Vilenkin:1998rp}).  Quantum mechanical
evolution should then take the initial wave function and evolve it up
to the present time.  The result will be a huge superposition of
possible outcomes.  From this wave function we need to determine what
we should expect to observe.  Presumably, most sectors of this wave
function will describe conditions not suited to the existence of
observers, so we must take into account anthropic considerations
\cite{Vilenkin:1995ua}.

With the treatment that I have advocated here, the procedure is
straightforward.  We look at all observers existing in all sectors of
the wave function, and assume that we are randomly situated among all
those observers, with probabilities given by the squared magnitudes of
the coefficients in the wave function.  This gives us the probability
distribution for what we should expect to observe.  If we have already
made some observations, and wish to predict the outcome of others,
then we should simply discard all observers whose observations conflict with
what we have seen and consider the distribution among those
observers remaining.

On the other hand, if we follow Leslie and Bostrom's analysis, then we
will have the problem of section \ref{sec:qm-interpretation} above,
and our conclusions will depend on our interpretation of quantum
mechanics.  If we use the Copenhagen interpretation, then sectors of
the wave function should be interpreted as classical probabilistic
alternatives.  Then, if we follow Leslie and Bostrom, we should expect
that we are typical observers in a typical sector of the wave
function, which is not the same thing as the above.  Sectors with
small numbers of observers are overrepresented in this procedure.  If,
on the other hand, we use the multiple worlds interpretation, then all
possible observers actually exist, and we should find ourselves to be
typical among them.  The fact that Leslie and Bostrom's treatment
yields ambiguous predictions argues against its use in evaluating
theories of quantum cosmology.

\section{Conclusion}
When you learn new information, you should update the probabilities you
assign to various hypotheses, based on the new information.  You should
now favor hypotheses that made the new data seem likely over those
that made it seem unusual.  Thus whatever chance you assign to the
possibility that the human race will end fairly shortly, you should
increase it when you take into account the position in which you find
yourself.  Knowing that you are among the first 70 billion people to
exist gives you reason to prefer theories in which the total number of
people ever to exist is not much larger than 70 billion.  This is the
Doomsday Argument \cite{Carter:original,Leslie:book,Bostrom:thesis}.

On the other hand, I have argued here that when you take into account
the fact that you exist at all, you should update your probabilities
in precisely the inverse manner, finding it more likely to be in a
race with a larger total number of individuals.  This effect follows
if one considers the case where there might be a large number or might
be a small number of people as analogous to the case where there are a
large number of people of one kind and a small number of another.
This effect exactly cancels the effect of taking into account your
position in the species.

The result of including both these effects is the same as the result
of including neither.  You judge the probability of early doom to be
whatever you judge it to be from considering the actual processes that
might put an end to humanity.  You get no other information from
arguments about your existence and when you were born.  This agrees with
our intuition that the chance of an event (e.g., the earth being hit by
an asteroid) should not depend on the event's consequences (e.g.,
humanity being wiped out).

In the case of Gott's argument, the situation is clear.  One must
include the fact that longer-lived phenomena are more likely to be
currently ongoing to get sensible predictions
\cite{Buch:letter,Caves:Gott}.  Once one does so, Gott's argument does
not change the prior probability distribution for the longevity of a
phenomenon.

In the case of the Carter-Leslie argument, the situation is not as
clear as that.  However, I have given a number of arguments that
we should in fact treat universes with longer-lived human races as
more likely for that reason, until we take account of our position in
the species.  With this effect included, there is no force in the
Doomsday Argument.

Nevertheless, I would not want anyone to become unconcerned with the future
of humanity as a result of the present paper.  Serious dangers do face
us, and we should work to minimize them, even if they are no larger
than one would at first think.

\section{Acknowledgments}

I would like to thank Alex Vilenkin for stimulating my interest in
this problem, and Judy Anderson, J. J. Blanco-Pillado, Allen Everett,
Larry Ford, Jaume Garriga, Xavier Siemens, Alex Vilenkin, and Valerie
White for useful conversations.  This work was supported in part by
grants from the John Templeton foundation and the National Science
Foundation.

%\bibliographystyle{unsrt}
%\bibliography{anthropic}

\end{document}